\documentclass[aps,prd,twocolumn,showpacs,superscriptaddress,groupedaddress, nofootinbib]{revtex4} 

\usepackage[utf8]{inputenc}
\usepackage{graphicx}  
\usepackage{dcolumn}   
\usepackage{bm}        
\usepackage{amssymb}   
\usepackage[colorlinks=true, linkcolor=blue, citecolor=blue, urlcolor=blue]{hyperref}
\usepackage{xcolor}
\usepackage{array}
\usepackage{booktabs} 
\usepackage{appendix}

\newcommand{\az}{a_{{\rm pec},\parallel}}
\usepackage{color}

\newcommand{\Pa}{P_{a_{\parallel}a_{\parallel}}}
\newcommand{\YZ}[1]{{\color{black} #1}}

\begin{document}

\title{Testing the equivalence principle on cosmological scales using peculiar acceleration power spectrum}

\author{Guoyuan Lu}
\affiliation{School of Physics and Astronomy, Sun Yat-Sen University, Zhuhai 519082, China}
\affiliation{CSST Science Center for the Guangdong–Hong Kong–Macau Greater Bay Area, SYSU, Zhuhai 519082, People’s Republic of China}

\author{Yi Zheng}
\email{zhengyi27@mail.sysu.edu.cn}
\affiliation{School of Physics and Astronomy, Sun Yat-Sen University, Zhuhai 519082, China}
\affiliation{CSST Science Center for the Guangdong–Hong Kong–Macau Greater Bay Area, SYSU, Zhuhai 519082, People’s Republic of China}

\author{Le Zhang}
 \email{zhangle7@mail.sysu.edu.cn}
 \affiliation{School of Physics and Astronomy, Sun Yat-Sen University, Zhuhai 519082, China}
 \affiliation{CSST Science Center for the Guangdong–Hong Kong–Macau Greater Bay Area, SYSU, Zhuhai 519082, People’s Republic of China}

\author{Xiaodong Li}
\email{lixiaod25@mail.sysu.edu.cn}
\affiliation{School of Physics and Astronomy, Sun Yat-Sen University, Zhuhai 519082, China}
\affiliation{CSST Science Center for the Guangdong–Hong Kong–Macau Greater Bay Area, SYSU, Zhuhai 519082, People’s Republic of China}

\author{Jiacheng Ding}
\email{dingjch@shao.ac.cn}
\affiliation{Shanghai Astronomical Observatory, Chinese Academy of Sciences, No. 80 Nandan Road, Shanghai
200030, China}

\author{Kwan Chuen Chan}
 \email{chankc@mail.sysu.edu.cn}
 \affiliation{School of Physics and Astronomy, Sun Yat-Sen University, Zhuhai 519082, China}
 \affiliation{CSST Science Center for the Guangdong–Hong Kong–Macau Greater Bay Area, SYSU, Zhuhai 519082, People’s Republic of China}


\date{\today}

\begin{abstract}
While the (weak) equivalence principle (EP) has been rigorously tested within the solar system, its validity on cosmological scales, particularly in the context of dark matter and dark energy, remains uncertain. In this study, we propose a novel method to test EP on cosmological scales by measuring the peculiar acceleration power spectrum of galaxies using the redshift drift technique.
We develop an EP estimator, $E_{\rm ep}$, to evaluate the consistency of the peculiar acceleration power spectrum across different tracers. By calculating the ratio of the peculiar acceleration power spectra of tracers, the ensemble average of $E_{\rm ep}$
is expected to be unity if EP holds on cosmological scales for these tracers. We validate this estimator using N-body simulations, focusing on four redshift bins with $z\leq 1.5$ and scales of $k$ in the range of $0.007$ and $0.2$ $h/\rm Mpc$. \YZ{By fitting a single parameter $\delta_{\rm ep}$ across redshifts, we find that DM particle mocks without EP violation yield $\delta_{\rm ep}$ consistent with zero under the small redshift measurement uncertainty case, while the large redshift uncertainty case slightly induces biases at low redshifts. In addition, when using DM halo mocks with controlled EP violations, no-violation and mild-violation cases show no significant detection, while moderate and strong violations produce statistically significant $\delta_{\rm ep}$ values and high $\chi^2$, especially at low redshifts, confirming the estimator’s sensitivity.} Taking advantage of advanced observing capabilities, such as next-generation facilities that extend beyond the Square Kilometer Array, the proposed method offers a promising approach for future cosmological tests of EP.

\end{abstract}
\keywords{methods: data analysis, numerical --- cosmology: large-scale structure of Universe, theory}

\maketitle

\section{Introduction}
\label{sec:intro}
The celebrated general theory of relativity is undoubtedly founded upon the equivalence principle (EP), which represents one of its cornerstones. EP states that all nongravitational phenomena are locally unaffected by gravity, provided that the experiments are performed in a freely falling frame. A specific consequence of this principle is that all objects (including light) fall in accordance with the same laws. Thus far, EP has been subjected to rigorous testing within the solar system~\citep{Will:2014kxa,Touboul:2017grn}. However, its applicability on cosmological scales when considering dark matter (DM) and dark energy remains uncertain. In essence, the large-scale structure of the Universe contains valuable information about EP, which can be used to directly evaluate its validity. To date, only indirect methods have been proposed to test EP~\citep{Kesden2006a,Kesden2006b,Kehagias:2013rpa, Creminelli:2013nua,Bonvin:2018ckp, Umeh:2020cag}. In this study, we propose a direct test of (weak) EP via the density-weighted peculiar acceleration power spectrum, which can be quantified through the redshift drift technique. 

The concept of redshift drift -- the change in the redshift of objects following cosmological expansion over time -- has been a subject of study for decades. The theory was initially proposed by~\cite{Sandage1962,1962ApJ...136..334M} and the concept was reinitiated approximately two decades ago by~\cite{1998ApJ...499L.111L}. Recently, drift motions in perturbed FRW universes, and a scale-dependent dipolar modulation of the acceleration of the expansion rate inferred from observations of objects within the bulk flow have been investigated by~\cite{2008PhLB..660...81A,2010MNRAS.405..503T,2019A&A...631L..13C}.

The measurement of redshift drift plays an essential role in real-time cosmology, as it is equivalent to observing the expansion of the Universe in real time. Moreover, the redshift drift is a model-independent method of probing the expansion of the Universe, thereby eliminating assumptions about geometry, clustering, or gravitational behavior. One of the key challenges in measuring redshift drifts arises from the vast difference between the time scales relevant to cosmology and human observation, requiring extremely precise measurements. The primary sources of uncertainty in redshift drift measurements arise from shot noise and redshift measurement precision. Addressing these challenges requires both a high number density of galaxies and spectroscopy with ultrahigh spectral resolution.

Recently, there have been conceptual discussions regarding potential measurement techniques and different cosmological observables of redshift drift~\citep{2007ApJ...671.1075L,Uzan:2007tc,Uzan:2008qp,2012IJMPD..2142017S,Neben:2012wc,Kim:2014uha,Koksbang:2015ctu,Martins:2016bbi}. Although the sensitivity of current facilities is still several orders of magnitude lower than that of the expected signal~\citep{2012ApJ...761L..26D}, it is promising to measure redshift drift as future facilities become available~\citep{2019MNRAS.488.3607A}.  For example, the Extremely Large Telescope (ELT), with its unique and highly advanced spectrographic capabilities~\citep{2008MNRAS.386.1192L, Balbi:2007fx, Corasaniti:2007bg, Lake:2007pf, Moraes:2011vq, Geng:2014ypa}, will facilitate a significant advancement in our understanding of cosmology by enabling the breaking of degeneracies between cosmological parameters. As recently observed by~\cite{Klockner:2015rqa,KangJG2025}, low redshift measurements are, in principle, feasible with the Square Kilometre Array (SKA). Similarly, ~\cite{Yu:2013bia} proposed that redshift drift measurements at intermediate redshifts could be conducted using intensity mapping techniques~\citep{2009astro2010S.234P, 2012IJMPS..12..256C}, such as those employed by the Canadian Hydrogen Intensity Mapping Experiment (CHIME)~\citep{2014SPIE.9145E..22B}. This analysis~\citep{Neben:2012wc} should also be applicable to the Hydrogen Intensity and Real-time Analysis eXperiment (HIRAX).

To rigorously test whether EP holds on cosmological scales, it is essential to examine the statistical internal consistency of the peculiar acceleration power spectrum of galaxies. This is a logical consequence of the EP test. As a first step, we propose in this paper an EP estimator to evaluate the principle and perform tests on two different mock datasets: DM particles and DM halos. In this study, we explore the potential of radio observations, focusing on an HI galaxy survey beyond the capabilities of the SKA. This survey is expected to cover $10^8$ HI galaxies at $z \leq 1.5$~\cite{Klockner:2015rqa,KangJG2025}. Moreover, we make a forward-looking assumption that sufficiently long integration times will enable redshift measurements with a precision of $\sim 1~{\rm cm/s}$. This precision is significantly higher than the expected capabilities of the SKA redshift drift experiment~\citep{Klockner:2015rqa}.

In the following, we will present a theoretical methodology for statistical measurements of the peculiar acceleration of galaxies, which will then be employed to quantify the EP test using mock data.

\section{Methodology}
\label{sec:method}

The galaxy acceleration along the line of sight (LOS), analogous to galaxy velocity, can be decomposed into two distinct components: the background cosmic acceleration, $a_{\rm cos}$, and a perturbed component, namely the peculiar acceleration $a_{\rm pec,\parallel}$. This decomposition can be expressed as  
\begin{equation}\label{eq:a}
    a_{\|} \equiv c\dot{z} = a_{\rm cos}+a_{\rm pec,\parallel}\,, 
\end{equation}
where $c$ is the speed of light. As initially demonstrated by~\citep{Sandage1962}, in a homogeneous and isotropic spacetime, the first term  is directly linked to the Hubble function through the relation:
\begin{equation}
    \label{eq:a_cos}
    a_{\rm cos} = c(1+z)H_0-cH(z)\,,
\end{equation}
where $H_0$ represents the present-day value of the Hubble parameter, $H(z)$. \YZ{This term forms the basis for the redshift drift technique, which is proposed to directly measure the real-time cosmic expansion~\cite{Sandage1962,1962ApJ...136..334M,1998ApJ...499L.111L}.} The second term of Eq.~\ref{eq:a}
accounts for the effect of local gravitational interactions, which cause the object to accelerate relative to the Hubble flow. This effect is determined by the gradient of the gravitational potential ($\phi$):
\begin{equation}
\label{eq:a_pec}
a_{\rm pec,\parallel}(\bm{x}, z) \equiv -\frac{\nabla \phi}{a}\cdot\hat{z}\,,
\end{equation}
where the unit vector $\hat{z}$ represents the LOS direction, and the gradient is computed in comoving coordinates.  

Note that when applying the redshift drift technique to detect peculiar accelerations, the full expression for the time derivative of galaxy redshift, up to first order in metric perturbations and in $v_{\rm pec}/c$, was derived in~\cite{Uzan:2008qp}. In writing Eq.~\ref{eq:a_pec}, we have made the following simplifications: i) we neglect all perturbative terms related to the local observer (us), as these terms vanish in the power spectrum estimator proposed in this study; ii) we omit two source-related terms: one analogous to the integrated Sachs-Wolfe (ISW) effect and another proportional to the peculiar velocity $v_{\rm pec}$. These terms are subdominant compared to the peculiar acceleration term, as demonstrated in~\cite{Uzan:2008qp}. \YZ{Furthermore, our analysis is conducted in the 3D redshift space under the plane-parallel limit approximation, though similar conclusions can be extended to 2D angular statistics at linear scales.}

\subsection{Estimator}
\label{subsec:power spectrum theory}

To test EP, we propose measuring the density-weighted peculiar acceleration power spectrum of galaxies using the redshift drift method. The corresponding acceleration power spectrum, $P_{a_{\parallel}a_{\parallel}}$, is expressed as follows,
\begin{eqnarray}\label{eq:pk_a}
    &&(2\pi)^3\delta^D(\bm{k}+\bm{k}^\prime)P_{a_{\parallel}a_{\parallel}}(\bm{k}) \nonumber \\
    &=& \left<(1+\delta_g)a_{\parallel}(1+\delta_g')a_{\parallel}'\right>\nonumber\\
    &=& \left<(1+\delta_g)\az(1+\delta_g')\az'\right>\nonumber\\   &\approx&\left<\az\az'\right>\,.
\end{eqnarray}
where, in the second equality, the constant term $a_{\rm cos}$ contributes only to the $k=0$ mode and is therefore ignored. 
This is because the $k=0$ mode represents the spatially uniform component of the field, which does not carry information about the spatial structure or clustering of galaxies. In the last equality, we restrict our analysis to linear scales and neglect the second-order term $\left\langle\delta_g a_{\rm pec,\|}\right\rangle$ \YZ{where the galaxy density fluctuation $\delta_g$ is much smaller than unity. In Appendix~\ref{app:validation}, we present the theoretical formulation of the linear $P_{a_{\parallel}a_{\parallel}}$ and compare it with simulation measurements to test its validity~\footnote{How the violation of this linear approximation will effect the accuracy of the estimator is tested in Sect.~\ref{sec:res}, where we demonstrate that even its breakdown due to nonlinearity on small scales does not significantly affect our results.}.} 


Consequently, $P_{a_{\parallel}a_{\parallel}}$ is unaffected by the galaxy bias at linear scales.  Based on this, we construct the following estimator, denoted as $E_{\rm ep}$, to test EP:
\begin{equation}\label{eq:E_ep}
    E_{\rm ep}(\bm{k} ) = \frac{P^i_{a_{\parallel}a_{\parallel}}(\bm{k})}{P^j_{a_{\parallel}a_{\parallel}}(\bm{k})}\,. 
\end{equation}
Here, the indices $i$ and $j$ represent distinct sets of a given acceleration tracer.
The primary advantage of this estimator lies in its ability to eliminate cosmic variance through the use of a ratio. By constructing the estimator as a ratio of two correlated quantities, the fluctuations contributed by cosmic variance exactly cancel out. This cancellation enables a more precise and reliable test of EP, significantly enhancing the robustness of the results.


If EP holds on cosmological scales for such tracers, the ensemble average of $E_{\rm ep}$ is expected to be unity. This expectation arises because the EP implies that all tracers should respond to gravitational forces in the same way, leading to a consistent relationship between their acceleration fields. Deviations from unity in the ensemble average of $E_{\rm ep}$ would indicate a violation of the EP or the presence of systematic effects.

\subsection{Simulation}
\label{subsec:mock}

The simulation tests in this work are performed in a N-body simulation run by the {\tt Gadget4}~\citep{Gadget4} code, adopting a $\Lambda$CDM cosmology with parameters: $\sigma_{8}=0.8228$, $n_{s}=0.96$, $\Omega_{m}=0.307115$, $\Omega_{b}=0.049$, and $h=0.68$, compatible with the Planck 2018 measurements within the $2\sigma$ level~\citep{Planck2018}. The simulation employs a box size of $L_{\rm box}=1024~{\rm Mpc}/h$ and a total of $N_{\rm p}=1024^3$ particles.  

\YZ{To validate the robustness of the estimator $E_{\rm ep}$ on the EP test, we examine three key factors that may affect its measurement: (i) the tracer number density, which influences the shot noise term in Eq.~\ref{eq:pk_an}; (ii) uncertainties in redshift measurements; and (iii) deviations from the linearity assumption underlying Eq.~\ref{eq:E_ep}. The first two factors mainly contribute to the statistical uncertainty in the $E_{\rm ep}$ measurement, while the third introduces a potential
systematic bias in the EP test.}

\YZ{Ideally, the test should be performed on a simulation combining sufficiently large volume and high mass resolution to both resolve small-mass halos and reproduce the tracer number density expected in upcoming HI surveys like SKA. Due to the limited volume and resolution of our current simulation, we instead employ a two-pronged validation approach: one test utilizes DM particles as tracers to minimize shot noise and isolate the impact of redshift measurement uncertainty on the $E_{\rm ep}$ detection, while the other uses DM halos to assess how shot noise contributes to the statistical uncertainty in the $E_{\rm ep}$ measurement, and how the nonlinear evolution of structure formation may introduce systematic bias in the interpretation of $E_{\rm ep}$}

\YZ{For the \textbf{DM particle test}, we treat individual DM particles as acceleration tracers and introduce random redshift measurement errors to mimic observational uncertainties. Thanks to the large number of particles, shot noise is well characterized and does not significantly bias the results. Furthermore, since both tracer populations in this test are drawn from the same underlying DM distribution, the expectation value of $E_{\rm ep}$ is analytically known to be unity and independent of wavenumber $k$. Therefore, this test is free from biases caused by the breakdown of the linear approximation in Eq.~\ref{eq:E_ep}.
}

\YZ{For the \textbf{halo-based test}, we use DM halos as tracers and assume an idealized scenario without redshift errors. This choice is motivated by the fact that redshift uncertainties are inherently significant for the EP test, and including them in a low-number-density sample would introduce substantial noise, severely degrading the precision of the measured acceleration power spectrum. In this test, we focus on the impact of shot noise and the potential breakdown of linearity on small scales. Together, these two complementary setups allow us to isolate and evaluate the effects of tracer density, redshift errors, and nonlinearity on the robustness of the EP estimator.}

\YZ{
Beyond quantifying the impact of different sources of uncertainty, we further divide our validation strategy into two physical scenarios based on whether the input tracer samples contain intrinsic EP violation. In the first scenario, the samples are constructed to obey EP, allowing us to evaluate how tracer number density, redshift uncertainties, and nonlinearity affect the signal-to-noise ratio of the $E_{\rm ep}$ measurement. In the second scenario, we introduce artificial EP violations into the halo catalogs by modifying the acceleration field in a controlled manner. This enables a direct assessment of the sensitivity and robustness of the $E_{\rm ep}$ estimator in detecting potential deviations from EP.
}

\subsubsection{DM particle mock generation}
\label{subsec:DM mock}

\YZ{We first test EP violation using the DM particle mock. To calculate the estimator $E_{\rm ep}$, which is defined as the ratio of power spectra from two distinct datasets, we proceed as follows. For each redshift snapshot, we randomly select $10^8$ particles from the full set of $1024^3$ DM particles in the simulation to form the first dataset. These selected particles are then excluded from the full set, and another $10^8$ particles are randomly chosen from the remaining particles to construct the second dataset. 

To mimic realistic observational uncertainties, we simulate redshift measurement errors by adding independent Gaussian noise to the redshift of each particle in both datasets.} The noise was generated with a mean of zero and a standard deviation of $\sigma_{z_{\parallel}} = f \sigma_{\rm SKA}/\sqrt{t_{\rm obs}}$, where $\sigma_{\rm SKA} = 10~{\rm m/s}$~\citep{Klockner:2015rqa}  represents the typical redshift measurement uncertainty expected from SKA observations (an  integration time with $\mathcal{O}(1)-\mathcal{O}(10)$ hours per pointing), and $t_{\rm obs}=12$ years. Here the parameter $f$ is a scaling factor applied to $\sigma_{\rm SKA}$. In this study, we consider two noise levels to assess their impact on SKA observation accuracy: \YZ{a \textbf{low-noise case} with $f = 0.001$, and a \textbf{high-noise case} with $f= 0.002$.} The latter corresponds to a doubled noise amplitude, resulting in four times higher noise power in the power spectrum.



\subsubsection{DM Halo Mock generation}
\label{subsec:halomock}

\begin{table}[t!]
\centering
\renewcommand{\arraystretch}{1.2} 
\setlength{\tabcolsep}{6pt} 
\begin{tabular}{c|c|r}
\hline
\hline
$z$ & $N_h$ (small/large) & $M_h[10^{12}~M_\odot/h$] (small/large)\\
\hline
1.5 & $360821/ 1082464$ &  $[2.73-3.41]/[3.41-847.050] $ \\
1.0 &  $ 466526/1399581$ & $ [2.73-3.49]/[3.49-1387.33] $ \\
0.5 &  $536364/ 1609094$ &  $ [2.73-3.58]/[3.58-2421.35] $ \\
0.0 &  $550123/1650372$ &  $[2.73-3.66]/[3.66-5085.91]$ \\
\hline
\hline
\end{tabular}
\caption{Number of halos in the ``small-mass'' and ``large-mass'' groups, along with their corresponding mass ranges, are summarized at different redshifts.}
\label{tab:halos}
\end{table}

In addition to testing EP using DM particle simulation data, we also conduct the second test by comparing the acceleration power spectra of DM halos with different masses. In the simulation, DM halos are identified using the group finder {\tt ROCKSTAR}~\citep{rockstar}. We selected all halos containing at least ten DM particles, corresponding to $M_h \geq 2.73 \times 10^{12}~M_\odot/h$.  These halos are subsequently divided into two mass bins, referred to as ``small-mass'' and ``large-mass'', with the former containing halos of smaller mass and the latter containing halos of larger mass (Tab.~\ref{tab:halos}). For each fixed redshift, the ratio of the number of small-mass halos to large-mass halos is maintained at approximately $1/3$. This ratio ensures that the shot noise contributions from the small and large mass bins are nearly equal, as confirmed by our measurements. 

\YZ{To assess the sensitivity of our estimator to violations of EP, we introduce artificial EP-breaking effects into the halo catalogs as follows. For each halo of mass $M$, we rescale its LOS acceleration by a mass-dependent factor $\lambda$:
\begin{equation}
\label{eq:lambda}
a_{\parallel, \mathrm{art}} = \lambda a_{\parallel}, \quad \text{with} \quad \lambda = 1+ \alpha \left( \frac{M}{M_\ast} \right)^\beta\,.
\end{equation}
This model is not derived from any specific underlying theory, but instead serves as a controlled, parametrized prescription to introduce EP violations as a function of halo mass. It is a purely phenomenological framework, designed to systematically probe the sensitivity of the estimator to such violations. When $\alpha = 0$, the EP is exactly preserved. Nonzero values of $\alpha$ introduce controlled deviations from EP.  In this study, the characteristic mass scale is fixed at $M_* = 10^{12} M_\odot$, as it represents the characteristic mass of typical galaxies.

We fix $\alpha = 0.1$ and vary $\beta$ to control the mass dependence of the EP violation. Within this phenomenological framework, we consider representative values of $\beta = 0.1$, $0.3$, and $0.5$ to assess how different scalings of the EP-breaking term influence the response of our estimator. Under these parameter choices, the relative deviations of $a_{||}$ from $\lambda$ in the small-mass group reach approximately 10\%-15\% for $\beta = 0.1$ and $0.3$, and up to 18\% for $\beta = 0.5$. In contrast, for the large-mass group, the deviations are more significant--about 20\% for $\beta = 0.1$, 100\% for $\beta = 0.3$, and a factor of 3.9–8.1 for $\beta = 0.5$.
}

\subsection{Measurements of $\Pa$ and its covariance}
\label{subsec:Pa_measure}

\YZ{The method by which we measure the power spectrum of the LOS acceleration, $\Pa$, from both observational and simulation data is as follows. In observations, we employ the redshift drift technique to infer the LOS acceleration $a_{||}$ via the relation $\dot{z} = a_{||}/c$ for each galaxy. In simulations, we directly extract the acceleration of DM particles or halos from the simulation snapshots. In particular, the acceleration of a halo is that of the particle nearest to the halo center.}

\YZ{
The computation of $\Pa$ follows an approach analogous to the standard method used for estimating the power spectrum of the mass-weighted velocity field (i.e., the momentum field). Specifically, we proceed as follows:

\begin{enumerate}
    \item We define a cubical volume enclosing the full survey or simulation region, as described in Sect.~\ref{subsec:mock}. For simulations, this naturally corresponds to the simulation box itself. A regular grid is then constructed within this volume.
    
    \item Using the cloud-in-cell (CIC) method, we interpolate the mass-weighted acceleration of tracers (galaxies, halos, or DM particles) onto the grid to construct a continuous acceleration field. For simulation data, prior to this assignment, we convert tracer positions from real space to redshift space along a chosen LOS direction, i.e., the $x$-axis, as:
    \begin{equation}
        x_s = x_r + \frac{v_x}{aH(z)}\,.
    \end{equation}
    Here, $x_s$ and $x_r$ denote the redshift-space and real-space positions along LOS, respectively; $v_x$ is the LOS component of the proper peculiar velocity; $a$ is the scale factor; and $H(z)$ is the Hubble parameter at redshift $z$.

    \item We perform a fast Fourier transform (FFT) on the gridded, mass-weighted LOS acceleration field to obtain its Fourier components. The power spectrum $\Pa(k)$ is then computed by averaging the squared amplitude of the Fourier modes over spherical shells in $k$-space, i.e.,  
$\Pa(k) = \langle |\az(\mathbf{k})|^2 \rangle_{|\mathbf{k}| = k}$. This yields the isotropic (angle-averaged) component of the power spectrum. Accordingly, in this study we focus exclusively on the monopole of $\Pa$, leaving any anisotropic features for future investigation.

\item After measuring the power spectrum \( P_{a_\parallel a_\parallel} \), we apply corrections for systematic effects arising from shot noise and redshift measurement uncertainties. First, we subtract the Poisson shot noise contribution, estimated by $ \langle a^2_{\rm pec,\parallel}/n_h \rangle$, such as
\begin{eqnarray}\label{eq:pk_an}
&&(2\pi)^3\delta^D(\bm{k}+\bm{k}^\prime)\Pa (\bm{k}) \nonumber\\
&=&\left<\az\az'\right>-\frac{\left<\az^2\right>}{n_h}\,,
\end{eqnarray}
where $n_h$ represents the number density of the tracer (halos or particles in this study), and $\left<\az^2\right>$ is the dispersion of $\az$ for such tracer. For the DM mocks, we further correct for the additional noise induced by redshift measurement errors. This redshift uncertainty introduces Gaussian-distributed noise in the acceleration measurements, effectively contributing an extra shot-noise-like component to the power spectrum. We estimate the impact by assigning Gaussian acceleration noise (low or high) to $10^8$ randomly distributed particles, yielding a noise power spectrum with amplitude $\sim 10^{-16}\,({\rm Mpc}/h)^3\,{\rm cm}^2/{\rm s}^4$.
\end{enumerate}
}

\YZ{
The covariance matrix for the acceleration power spectrum or $E_{\rm ep}$ is  accurately estimated using the jackknife method~\citep{Norberg:2008tg} on a single mock dataset. Jackknife realizations are constructed by sequentially deleting one sample at a time and calculating the power spectrum from the remaining data. Specifically, using the delete-one jackknife approach, the simulation box is divided into $n_{s} = 64$ equal-volume sub-boxes. Each jackknife box contains $16^3$ cells on the CIC mesh. The covariance matrix for $E_{\rm ep}$ is estimated as
\begin{eqnarray}
\label{eq:cov}
C_{ij} = \frac{n_s - 1}{n_s} \sum_{l=1}^{n_s} && 
\left[ E_{\rm ep}^{[l]}(k_i) - \langle E_{\rm ep}(k_i) \rangle \right] \nonumber\\
&\times& \left[ E_{\rm ep}^{[l]}(k_j) - \langle E_{\rm ep}(k_j) \rangle \right]\,,
\end{eqnarray}
where $ E^{[l]}_{\rm ep}$ is the estimate of the $l$-th realization, and $\langle E_{\rm ep}\rangle$ is the mean value  from $n_{s}$ jackknife realizations. Analysis of the mock data reveals that the resulting covariance matrix is almost diagonal, with negligible correlations between different $k$-bins. 
}


\section{Results}\label{sec:res}

Here, we present the measured $E_{\rm ep}$ values for DM particle and halo mock data across four redshift bins of $z=[0.0, 0.5, 1.0, 1.5]$, with 10 $k$-bins ranging from $0.007$ to \YZ{$0.2~h/\rm Mpc$}. The $k$-bins are evenly spaced on a logarithmic scale. \YZ{For $k\gtrsim 0.2~h/\rm Mpc$, the measured power spectrum is roughly an order of magnitude lower than the shot noise level.} 
As a result, we restrict our analysis to scales below this threshold.

To quantitatively characterize the degree of deviation from the expected value, we propose a simple one-parameter model to fit the observed data $E^{\rm obs}_{\rm ep}(k)$ over $k$ bins. The parametrization is expressed as:
\begin{equation}\label{eq:fit} 
E^{\rm model}_{\rm ep}  = 1 + \delta_{\rm ep}\,,
\end{equation}
where $\delta_{\rm ep}$ represents the deviation from unity, which corresponds to the expected value if EP is perfectly satisfied. The corresponding $\chi^2$ is given by
\begin{equation}\label{eq:fit-chi} 
\chi^2= \sum_k \frac{\left[E^{\rm obs}_{\rm ep}(k)-E^{\rm model}_{\rm ep} \right]^2}{\sigma^2_k}\,,
\end{equation}
where correlations between different bins are neglected, as they are negligibly small. The variance $\sigma^2_k$ in the $k$-th bin is estimated using the jackknife method based on the simulations. Since the fit involves one free parameter, $\delta_{\rm ep}$, over 10 $k$-bins, the degrees of freedom are $N_{\rm dof} = 10 - 1 = 9$. The corresponding $2\sigma$ confidence interval for $N_{\rm dof} = 9$ corresponds to $\chi^2$ values between approximately 2.7 and 19.02.

\subsection{Results from DM Particle Mock}
\label{subsec:DM_results}

\begin{figure*}[t!]
    \centering
    \includegraphics[width=0.45\linewidth]{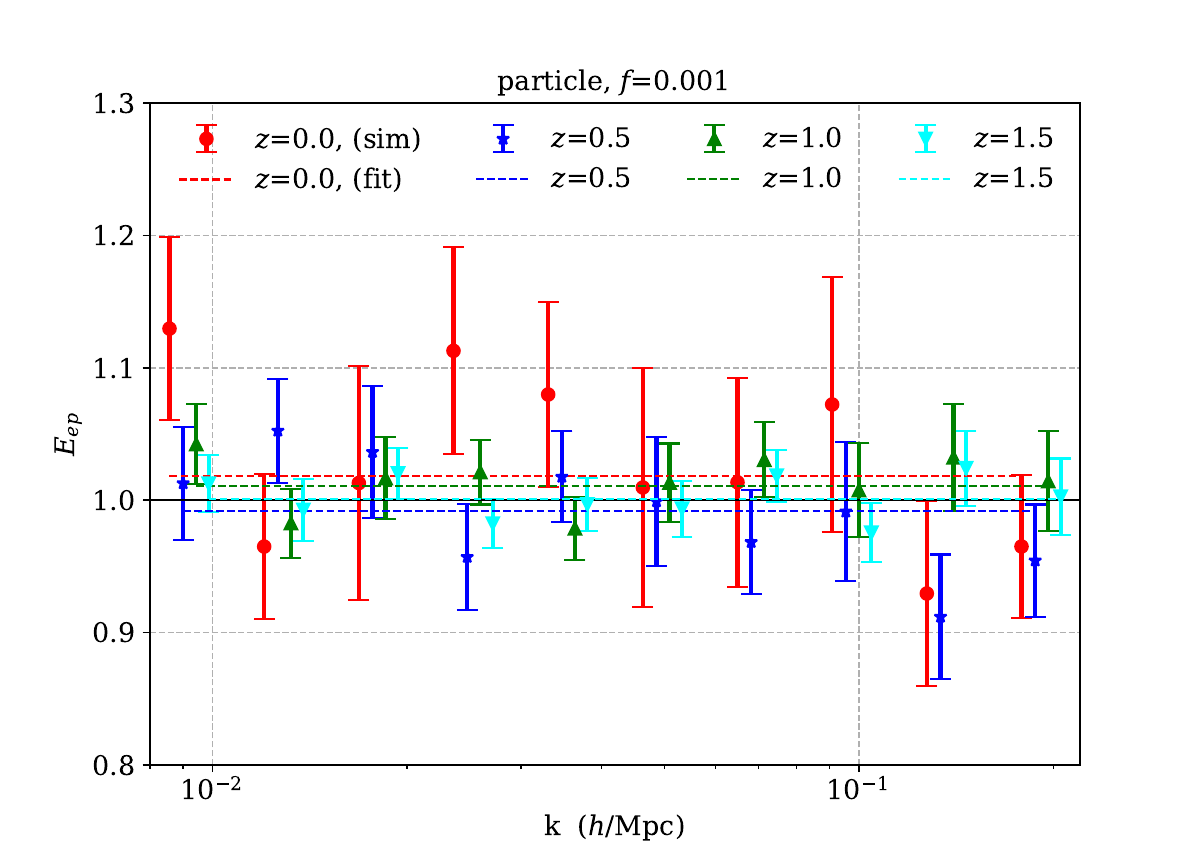}
    \includegraphics[width=0.45\linewidth]{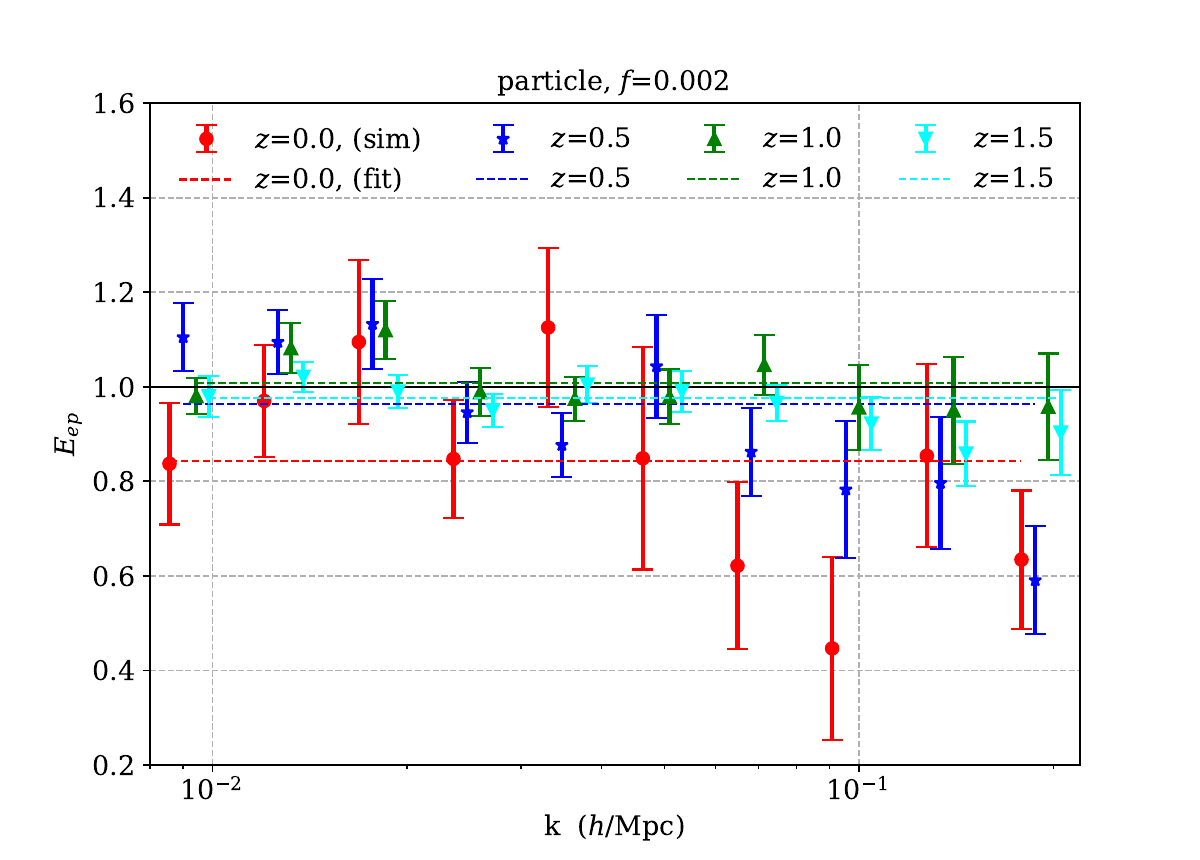}
    \caption{Measurement of $E_{\rm ep}$ for \textbf{DM particles} for testing EP. The ratio of acceleration power spectra for the two datasets--each containing $10^8$ particles--was computed after adding Gaussian redshift errors to each particle. These errors have zero mean and a standard deviation of $\sigma_{z_{\parallel}}$, corresponding to $f = 0.001$ (left) and $f = 0.002$ (right) for the low- and high-noise cases, respectively. Four redshift bins are considered, with $k$ values spanning the range $[0.007, 0.21]$ divided into 10 logarithmically spaced bins.  The black solid line represents the expected value $E_{\rm ep} = 1$ under the exact validity of EP. The error bars correspond to the \YZ{$1\sigma$} level, estimated from the jackknife scheme applied to the simulation box. \YZ{The fitted values of $E_{\rm ep}$ are also shown for comparison}. As shown, when $f=0.001$, the $E_{\rm ep}$ values remain consistent with unity within the $2\sigma$ confidence level across all $z$-bins. \YZ{When $f=0.002$, the $E_{\rm ep}$ values exhibit slightly larger deviations from unity and increased uncertainties, particularly at low redshifts}. Note that the bin centers have been slightly adjusted for clarity in the illustration. }
    \label{fig:ep-particle}
\end{figure*}

\begin{table}[!]
\centering
\renewcommand{\arraystretch}{1.2} 
\begin{tabular}{c|c|c|c}
\hline
\hline
$z$ (particle) & $\delta_{\rm ep}$ (\YZ{low/high}) & $\sigma_{\delta}$ (low/high) & $\chi^2$ (low/high) \\
\hline
0.0 & 0.0186/$-0.1563$ & 0.0225/0.0492 & 9.42/24.09 \\
0.5 & $-0.0084$/$-0.0364$ & 0.0136/0.0272 & 9.17/29.59 \\
1.0 & 0.0108/0.0083 & 0.0096/0.0184 & 6.45/8.01 \\
1.5 & 0.0013/$-0.0235$ & 0.0072/0.0135 & 5.39/11.08 \\
\hline
\hline
\end{tabular}

\caption{Best-fit values of $\delta_{\rm ep}$ and the corresponding $1\sigma$ statistical uncertainties, along with the resulting $\chi^2$ values for four $z$-bins, are presented for the \textbf{DM particle} simulation data. \YZ{To evaluate the impact of observational noise, the results compare the low- and high-noise cases across different SKA observational accuracies. }}
\label{tab:fit-dm}
\end{table}

Fig.~\ref{fig:ep-particle} presents the measured $E_{\rm ep}$ for DM particle mock data, \YZ{under the fact that EP holds in the underlying physical model. The primary aim here is to assess how redshift measurement uncertainties impact the signal-to-noise of the EP test.} 
Specifically, we compute the ratio of acceleration power spectra for two datasets (as described in Sect.~\ref{subsec:DM mock}). Two different levels of redshift measurement errors are introduced in the simulation particles: the left and right panels correspond to the low-noise case of $f=0.001$  and the high-noise case of $f=0.002$, respectively. As shown, the error bars in the right panel are visibly larger than those in the left panel, which is due to the relatively larger $f$ value. 

Based on the measured $E_{\rm ep}$, the model fitting results are summarized in Tab.~\ref{tab:fit-dm}, presenting the best-fit values of $\delta_{\rm ep}$, their associated $1\sigma$ statistical uncertainties, and the resulting $\chi^2$ values for the four $z$-bins.  As seen, the average error in the high-noise case is greater than that in the low-noise case by a factor of $1.56$, which is almost consistent with the expected scaling of $2$. This result demonstrates that the errors are significantly influenced by the redshift measurement error. 

\YZ{Specifically, in the low-noise case, the fitted values of $\delta_{\rm ep}$ remain close to zero across all redshift bins, with relatively small statistical uncertainties and moderate $\chi^2$ values. At $z = 0.0$, $\delta_{\rm ep} = 0.0186 \pm 0.0225$, and the corresponding $\chi^2$ is $9.42$, indicating a statistically consistent result with no significant deviation from the null hypothesis. Similar behavior is observed at higher redshifts: at $z = 0.5$, the best-fit value is $-0.0084 \pm 0.0136$ with $\chi^2 = 9.17$; at $z = 1.0$, $\delta_{\rm ep} = 0.0108 \pm 0.0096$ with $\chi^2 = 6.45$; and at $z = 1.5$, the value further reduces to $0.0013 \pm 0.0072$ with $\chi^2 = 5.39$. These results demonstrate the robustness of the $\delta_{\rm ep}$ estimator in the low-noise case, where statistical errors are well controlled and no significant bias is introduced. In summary, the $E_{\rm ep}$ values are generally consistent with unity within the $2\sigma$ level in most cases. Given the number of degrees of freedom $N_{\rm dof} = 9$, all $\chi^2$ values lie within the range $\chi^2 \in [0.51, 12.48]$, which is consistent with the expected $\chi^2$ distribution at the $2\sigma$ confidence level. These results validate the $E_{\rm ep}$ estimator as a robust and unbiased tool for testing the equivalence principle on linear cosmological scales, specifically over the range $k \in [0.007, 0.2]~h/\mathrm{Mpc}$.

In contrast, the high-noise case introduces larger deviations and uncertainties in the fitted $\delta_{\rm ep}$ values, as well as increased $\chi^2$ values indicating poorer fit quality. At $z = 0.0$, $\delta_{\rm ep}$ shifts to $-0.1563$ with a larger uncertainty of $0.0492$ and a much higher $\chi^2 = 24.09$. At $z = 0.5$, the estimate becomes more negative ($-0.0364$), and the $\chi^2$ rises significantly to $29.59$. These deviations suggest a noise-induced bias that compromises the reliability of the estimator. While the impact of noise is still visible at $z = 1.0$ and $z = 1.5$, its magnitude decreases: for instance, at $z = 1.0$, the best-fit value remains close to zero ($0.0083$), and $\chi^2$ increases modestly to $8.01$. At $z = 1.5$, $\delta_{\rm ep} = -0.0235$ with $\chi^2 = 11.08$.

Overall, these trends indicate that acceleration noise has a stronger impact at low redshift, leading to biased estimates and reduced statistical significance. In contrast, the low-noise case provides stable and unbiased estimates with tighter constraints. This highlights the critical importance of minimizing observational noise in future SKA surveys, especially for low-redshift measurements aiming to test EP with high precision.}

\subsection{Results from DM Halo Mock}\label{subsec:halo_results}
To evaluate the sensitivity of our estimator to violations of EP, we modified the LOS acceleration by a mass-dependent factor according to Eq.~\ref{eq:lambda}, thereby introducing artificial EP-violating effects into the halo catalogs.

\begin{figure*}[t!]
    \centering
    \includegraphics[width=0.45\linewidth]{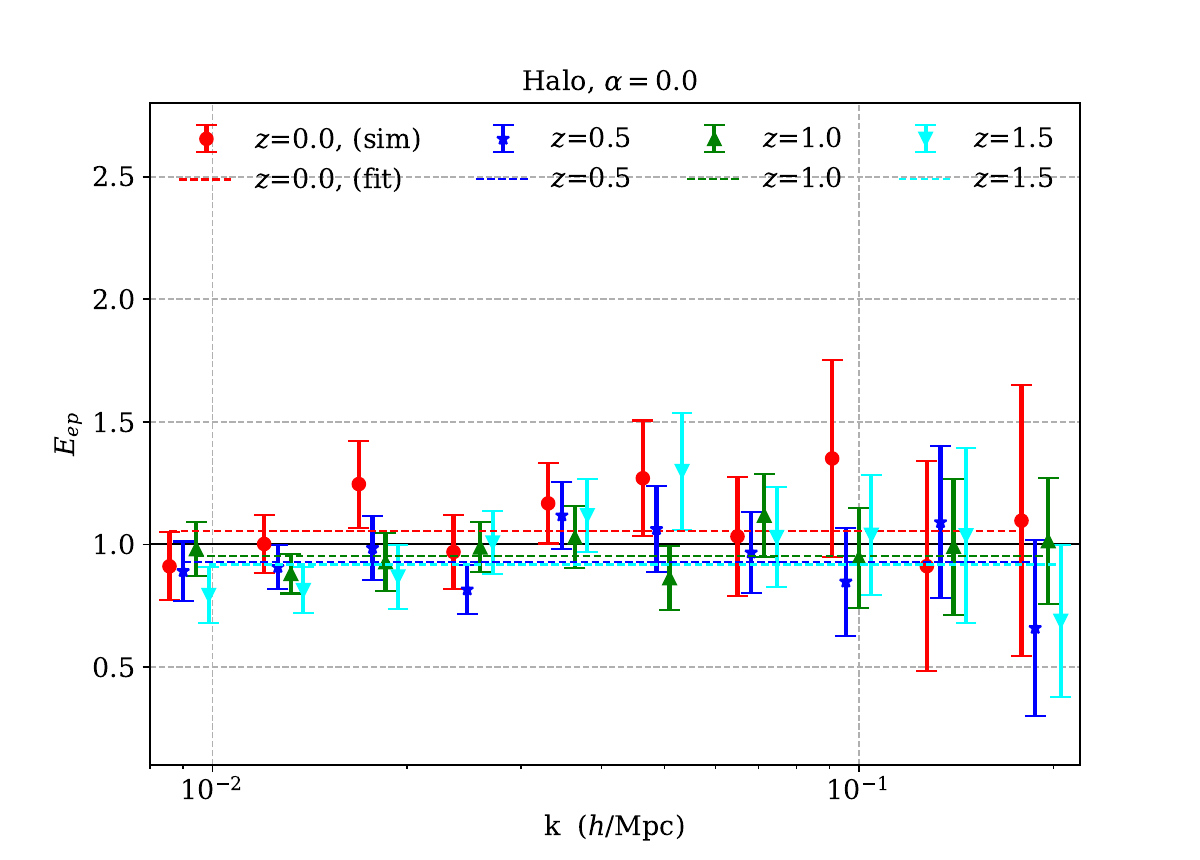}
    \includegraphics[width=0.45\linewidth]{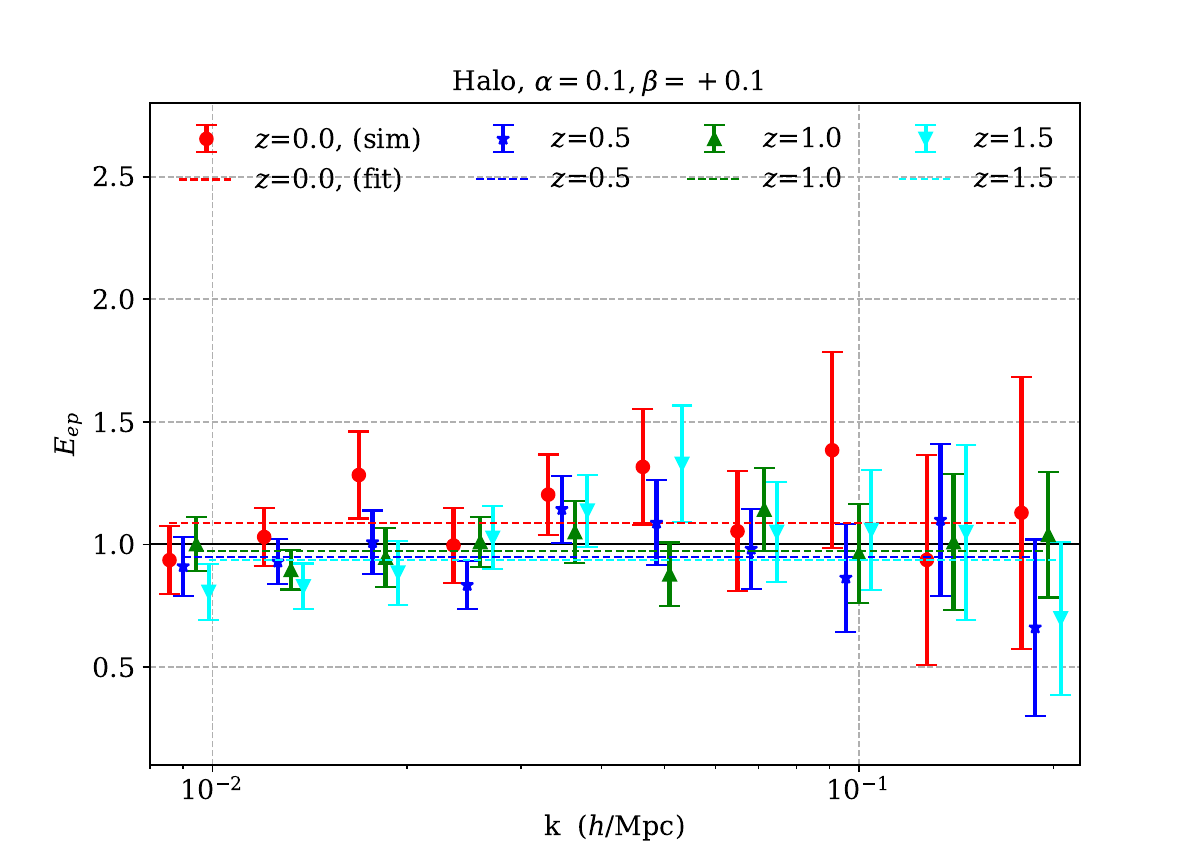}
    \includegraphics[width=0.45\linewidth]{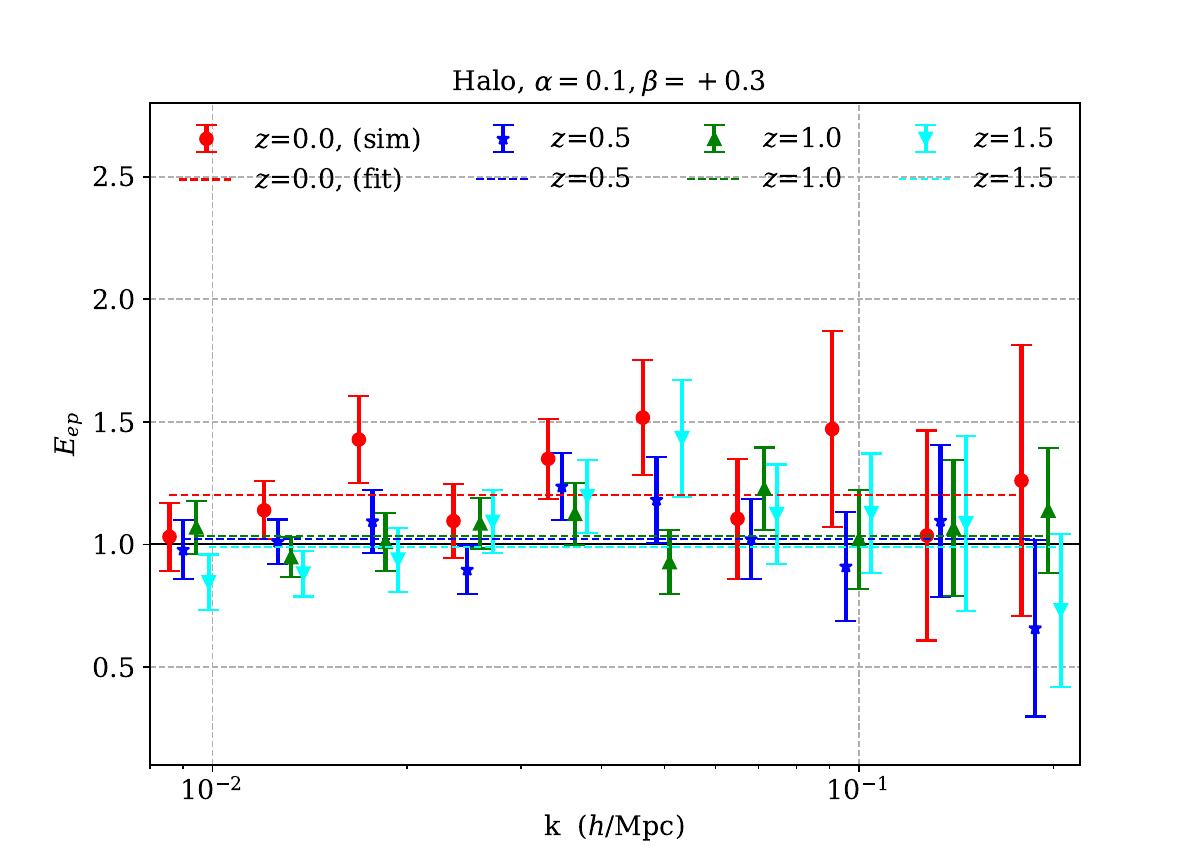}
    \includegraphics[width=0.45\linewidth]{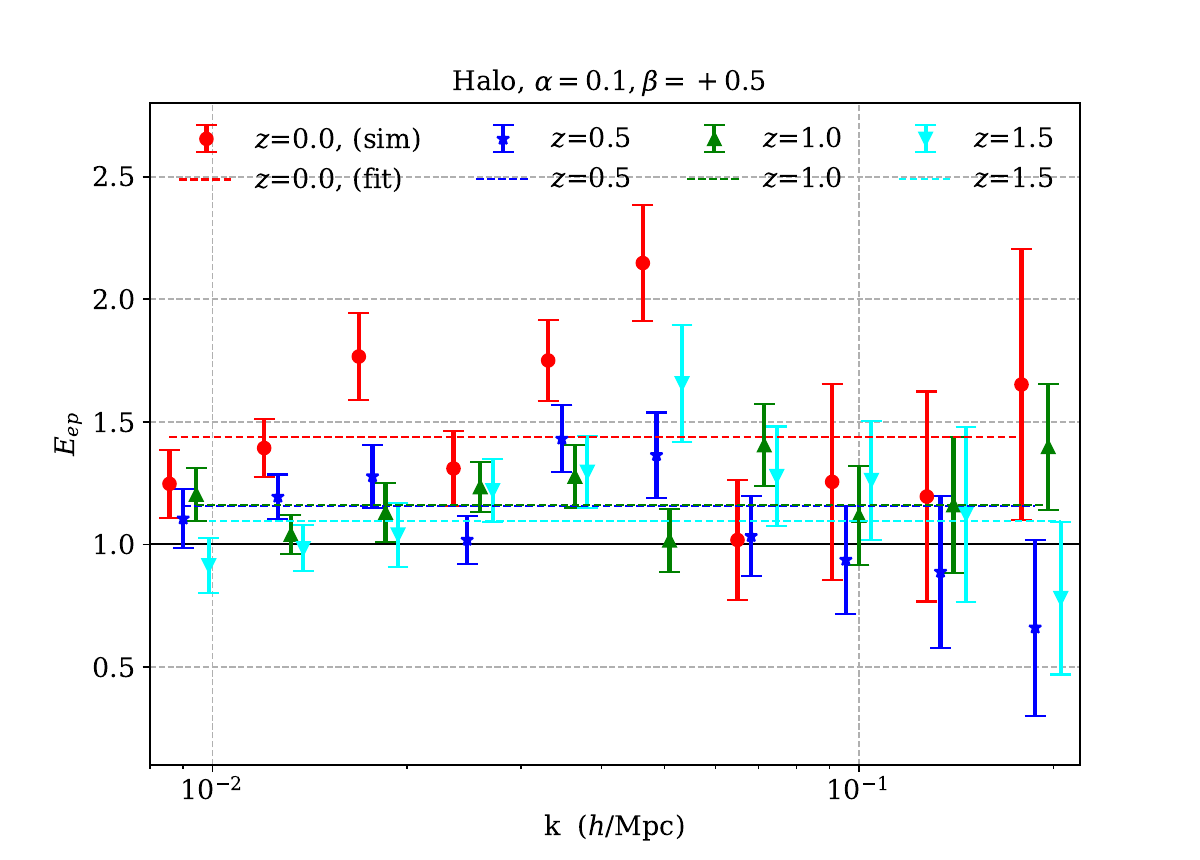}
    \caption{\YZ{Same as Fig.~\ref{fig:ep-particle}, but for the \textbf{DM halo} mock including cases with EP violation. The estimator $E_{\rm ep}$ is computed from the ratio of acceleration power spectra between the ``large-mass'' and ``small-mass'' halo catalogs. Shown are the estimated values of the EP violation parameter $\delta_{\rm ep}$ across different redshifts, for various levels of EP violation characterized by the parameters $(\alpha, \beta)$. The case with $(\alpha, \beta) = (0, 0)$ shows $\delta_{\rm ep} \approx 0$ across all redshifts, consistent with no violation. Mild violation scenarios such as $(0.1, 0.1)$ yield similar results, with $\delta_{\rm ep}$ remaining within $1\sigma$ uncertainties. In contrast, moderate $(0.1, 0.3)$ and strong $(0.1, 0.5)$ violations produce increasingly significant positive shifts in $\delta_{\rm ep}$, especially at lower redshifts, accompanied by rising $\chi^2$ values--indicating robust sensitivity of the estimator to EP-breaking signals.}}
    \label{fig:ep-halo}
\end{figure*}

\YZ{In Fig.~\ref{fig:ep-halo}, we present the measured values of the estimator $E_{\rm ep}$ using the DM halo mock data across four redshift bins, considering both EP-preserving and EP-violating scenarios. In the top-left panel, corresponding to the EP-preserving case with $\alpha = 0$ and $\beta = 0$, the measured $E_{\rm ep}$ values are generally consistent with the expected value of unity within the $2\sigma$ uncertainty range. As the wavenumber $k$ increases, the error bars become larger, primarily due to the increasing contribution of shot noise at smaller scales.} After fitting the model, we find that the best-fit $\delta_{\rm ep}$ values are consistent with zero within the $2\sigma$ level, indicating that deviations from unity are not statistically significant. Furthermore, the $\chi^2$ values fall within the expected range, supporting the validity of the proposed $E_{\rm ep}$ estimator for the EP test based on the halo simulation data. \YZ{In particular, as shown in Appendix~\ref{app:validation}, although the halo acceleration power spectrum deviates from the linear theory prediction at $k > 0.1~h/{\rm Mpc}$ due to nonlinear structure formation, the close agreement between $\Pa$ of low- and high-mass halos ensures that the estimator $E_{\rm ep}$ remains consistent with unity up to $k \sim 0.2~h/{\rm Mpc}$. Therefore, the nonlinearity of structure formation does not constitute a significant source of systematic error for our estimator in the relevant scale range.}

\YZ{For the rest of the panels in Fig.~\ref{fig:ep-halo}, we show the results for samples with artificial EP violation, implemented by varying $\beta = 0.1$, $0.3$, and $0.5$, as described in Eq.~\ref{eq:lambda}. As $\beta$ increases, the deviation of $E_{\rm ep}$ from unity becomes more pronounced. For instance, at redshift $z = 0$, the measured $E_{\rm ep}$ values at $k = 0.046~h/{\rm Mpc}$ are $1.23$, $1.31$, $1.51$, and $2.15$ for $\beta = 0$, $0.1$, $0.3$, and $0.5$, respectively, all with nearly identical uncertainties of $\pm 0.24$. Similar trends are observed at other redshifts, with the magnitude of the deviation increasing approximately proportionally with $\beta$ and becoming more detectable at larger scales (smaller $k$), where shot noise is comparatively lower.  These results quantitatively confirm that our estimator $E_{\rm ep}$ is capable of capturing EP-violating signals with high sensitivity over a broad range of redshifts.}

\begin{table*}[!]
\caption{\YZ{Summary of best-fit $\delta_{\rm ep}$ values with $1\sigma$ uncertainties $\sigma_\delta$ and corresponding $\chi^2$ statistics across four redshift bins, for the DM mock catalogs generated under different EP-violation parameters $(\alpha, \beta)$ indicated in the top row.}
}

\centering
\renewcommand{\arraystretch}{1.2} 
\setlength{\tabcolsep}{6pt} 
\begin{tabular}{cc}
\begin{tabular}{c|c|c|c}
\hline
\multicolumn{4}{c}{\boldmath{$\alpha=0$, $\beta=0$}} \\
\hline
$z$ & $\delta_{\rm ep}$ & $\sigma_{\delta}$ & $\chi^2$ \\
\hline
0.0 & 0.0542 & 0.0587 & 5.59 \\
0.5 & $-0.0708$ & 0.0442 & 7.72 \\
1.0 & $-0.0485$ & 0.0400 & 4.38 \\
1.5 & $-0.0823$ & 0.0476 & 11.63 \\
\hline
\hline
\end{tabular}
&
\begin{tabular}{c|c|c|c}
\hline
\multicolumn{4}{c}{\boldmath{$\alpha=0.1$, $\beta=0.1$}} \\
\hline
$z$ & $\delta_{\rm ep}$ & $\sigma_{\delta}$ & $\chi^2$ \\
\hline
0.0 & 0.0853 & 0.0589 & 7.20 \\
0.5 & $-0.0503$ & 0.0439 & 6.73 \\
1.0 & $-0.0290$ & 0.0402 & 3.69 \\
1.5 & $-0.0651$ & 0.0478 & 10.97 \\
\hline
\hline
\end{tabular}
\\[2.5em]
\begin{tabular}{c|c|c|c}
\multicolumn{4}{c}{\boldmath{$\alpha=0.1$, $\beta=0.3$}} \\
\hline
$z$ & $\delta_{\rm ep}$ & $\sigma_{\delta}$ & $\chi^2$ \\
\hline
0.0 & 0.1996 & 0.0586 & 18.80 \\
0.5 & 0.0209 & 0.0439 & 6.92 \\
1.0 & 0.0352 & 0.0401 & 5.00 \\
1.5 & $-0.0095$ & 0.0475 & 10.74 \\
\hline
\end{tabular}
&
\begin{tabular}{c|c|c|c}
\multicolumn{4}{c}{\boldmath{$\alpha=0.1$, $\beta=0.5$}} \\
\hline
$z$ & $\delta_{\rm ep}$ & $\sigma_{\delta}$ & $\chi^2$ \\
\hline
0.0 & 0.4366 & 0.0426 & 83.87 \\
0.5 & 0.1562 & 0.0437 & 25.37 \\
1.0 & 0.1597 & 0.0402 & 23.80 \\
1.5 & 0.0968 & 0.0477 & 18.88 \\
\hline
\end{tabular}
\end{tabular}
\label{tab:delta_results}
\end{table*}

\YZ{The results presented in Tab.~\ref{tab:delta_results} quantitatively illustrate the sensitivity of the $\delta_{\rm ep}$ estimator to different levels of EP violation, parametrized by $(\alpha, \beta)$. For the no-violation case with $\alpha=0$ and $\beta=0$, the best-fit $\delta_{\rm ep}$ values remain close to zero across all redshifts, ranging from approximately $-0.08$ to $0.05$, and are consistent with the $1\sigma$ uncertainties (about $0.04-0.06$). Corresponding $\chi^2$ values range from 4.38 to 11.63, indicating statistically acceptable fits. 

When a mild violation is introduced ($\alpha=0.1$, $\beta=0.1$), the results remain similar to the no-violation case ($\alpha=0$, $\beta=0$), with $\delta_{\rm ep}$ values exhibiting only slight shifts between $-0.065$ and $0.085$ and comparable uncertainties. The $\chi^2$ values show no significant increase, indicating that this small violation produces negligible deviation from the null hypothesis.  For moderate violation $(\alpha=0.1, \beta=0.3)$, the low-redshift bin ($z=0$) shows a substantial $\delta_{\rm ep} = 0.20 \pm 0.06$ with $\chi^2=18.80$, exceeding the $2\sigma$ threshold and indicating a potential EP violation detection. Higher redshifts exhibit smaller shifts and acceptable $\chi^2$ values.

In the strongest violation scenario $(\alpha=0.1, \beta=0.5)$, $\delta_{\rm ep}$ increases markedly, reaching $0.44 \pm 0.04$ at $z=0$ with an extreme $\chi^2 = 83.87$, far above the $2\sigma$ limit, strongly rejecting the null hypothesis.  The $\delta_{\rm ep}$ values remain significantly above zero at higher redshifts, measuring $0.156 \pm 0.043$ at $z=0.5$, $0.159 \pm 0.040$ at $z=1.0$, and $0.096 \pm 0.047$ at $z=1.5$, with corresponding high $\chi^2$ values that indicate notable deviations across these redshifts. Correspondingly, the $\chi^2$ values rise sharply up to 83.87 at $z=0$ and remain high at other redshifts, strongly rejecting the no-violation hypothesis. These quantitative trends demonstrate that the estimator robustly detects EP violations, with stronger violations yielding larger $\delta_{\rm ep}$ offsets.

Overall, the observed $\chi^2$ values demonstrate that our estimator reliably identifies EP violations, with stronger violations producing larger deviations in both $\delta_{\rm ep}$ and $\chi^2$, thus quantitatively confirming the estimator’s effectiveness.}

\section{Conclusion}
\label{sec:conclusion}
The (weak) EP has been extensively tested within the Solar System; however, its validity on cosmological scales remains an open question. In this study, we propose a novel approach to test EP on cosmological scales by measuring the peculiar acceleration power spectrum of galaxies through the redshift drift technique.

To achieve this, we develop an EP estimator, $E_{\rm ep}$, designed to evaluate the consistency of peculiar acceleration power spectra across different tracers.  If EP holds on cosmological scales for these tracers, the ensemble average of $E_{\rm ep}$ is expected to equal 1. As a first step toward testing EP on cosmological scales, we validate this estimator using N-body simulations. Applying this approach, we measure $E_{\rm ep}$ at four redshift bins ($z=[0, 0.5, 1, 1.5]$) on the linear scales of $k\in[0.007, 0.2]~h/\rm Mpc$.
This analysis is conducted using two distinct tests: i) DM particle mock data, incorporating redshift measurement errors, and ii) mock data for small-mass and large-mass DM halos.

\YZ{To quantitatively evaluate deviations of $E_{\rm ep}$ from unity, we fit a one-parameter model to obtain the best-fit $\delta_{\rm ep}$ values across different redshifts. For the DM particle mock assuming no EP violation, we computed the ratio of acceleration power spectra for two DM particle datasets with different redshift measurement errors, characterized by noise parameters $f=0.001$ (low-noise) and $f=0.002$ (high-noise). As expected, the high-noise case exhibits error bars larger by a factor of $\sim 1.56$, close to the theoretical scaling of 2, demonstrating the strong influence of redshift noise on uncertainties. In the low-noise case, the fitted $\delta_{\rm ep}$ values remain close to zero across all redshifts, consistent within the $2\sigma$ confidence interval. Similar results at higher $z$ confirm the estimator’s robustness and absence of bias in low-noise conditions. Conversely, in the high-noise case, $\delta_{\rm ep}$ deviates significantly from zero at low redshifts, indicating noise-induced bias and poorer fit quality. These deviations decrease at higher redshifts as the effect of noise becomes less significant.}

\YZ{Furthermore, for the DM halo mock, we considered various EP violation scenarios based on a phenomenological model where the violation strength is controlled by parameters $(\alpha, \beta)$. To quantitatively assess deviations of $E_{\rm ep}$ from unity, we fit the parameter $\delta_{\rm ep}$. For the no-violation case $(\alpha=0, \beta=0)$, $\delta_{\rm ep}$ values are consistent with zero within uncertainties of $\sim0.04$–$0.06$, and $\chi^2$ values fall within the $2\sigma$ confidence interval for $N_{\rm dof}=9$. Mild violation $(\alpha=0.1, \beta=0.1)$ produces small shifts in $\delta_{\rm ep}$ and similar $\chi^2$, indicating no significant detection. Moderate violation $(\alpha=0.1, \beta=0.3)$ shows a notable deviation at $z=0$, exceeding the $2\sigma$ limit. In the strongest case $(\alpha=0.1, \beta=0.5)$, $\delta_{\rm ep}$ reaches $0.44 \pm 0.04$ at $z=0$ with $\chi^2=83.87$, and remains significantly positive at higher redshifts with correspondingly high $\chi^2$, strongly rejecting the null hypothesis. These results demonstrate the estimator’s robust sensitivity to EP violations, with detection strength scaling with $\beta$.}

This study serves as a proof of concept for testing EP using N-body simulations. Since the simulations do not include any physical mechanisms that could induce EP violations, our measurements are expected to confirm the validity of $E_{\rm ep}\approx 1$ on large (linear) scales. This expectation is supported by our results, which show no significant deviation from unity when no artificial EP violation is introduced. \YZ{Furthermore, we have constructed mock catalogs with artificially introduced EP violations using DM halo samples, and demonstrated that our estimator $E_{\rm ep}$ is capable of detecting such deviations in a statistically significant manner. This confirms that $E_{\rm ep}$ is sensitive to physical violations of EP while remaining robust against shot noise and redshift-space distortions. In future studies, we will apply this estimator to real, high-precision observational data and further refine its application under more complex, observationally realistic conditions, as well as marginalizing over systematic effects to ensure reliable inference. Such efforts will open a new avenue for testing fundamental physics with large-scale structure surveys.}

{\it Acknowledgments:} We thank Volker Springel for useful discussions. We thank the anonymous referee for useful suggestions that significantly improve the quality of paper. This work is supported by the National SKA Program of China (2020SKA0110401, 2020SKA0110402, 2020SKA0110100), the National Key R\&D Program of China (2020YFC2201600, 2018YFA0404504, 2018YFA0404601), the National Science Foundation of China (12203107, 12473097, 12073088, 12373005, 12273121), the China Manned Space Project with No. CMS-CSST-2021 (A02, A03, B01), the Guangdong Basic and Applied Basic Research Foundation (2024A1515012309), and the Fundamental Research Funds for the Central Universities, Sun Yat-sen University (No. 24qnpy122). We also wish to acknowledge the Beijing Super Cloud Center (BSCC) and Beijing Beilong Super Cloud Computing Co., Ltd (\url{http://www.blsc.cn/}) for providing HPC resources that have significantly contributed to the research results presented in this paper.

\bibliography{mybib}{}

\appendix

\section{Validation of acceleration power spectrum}
\label{app:validation}

\begin{figure*}[t!]
    \centering
    \includegraphics[width=0.45\linewidth]{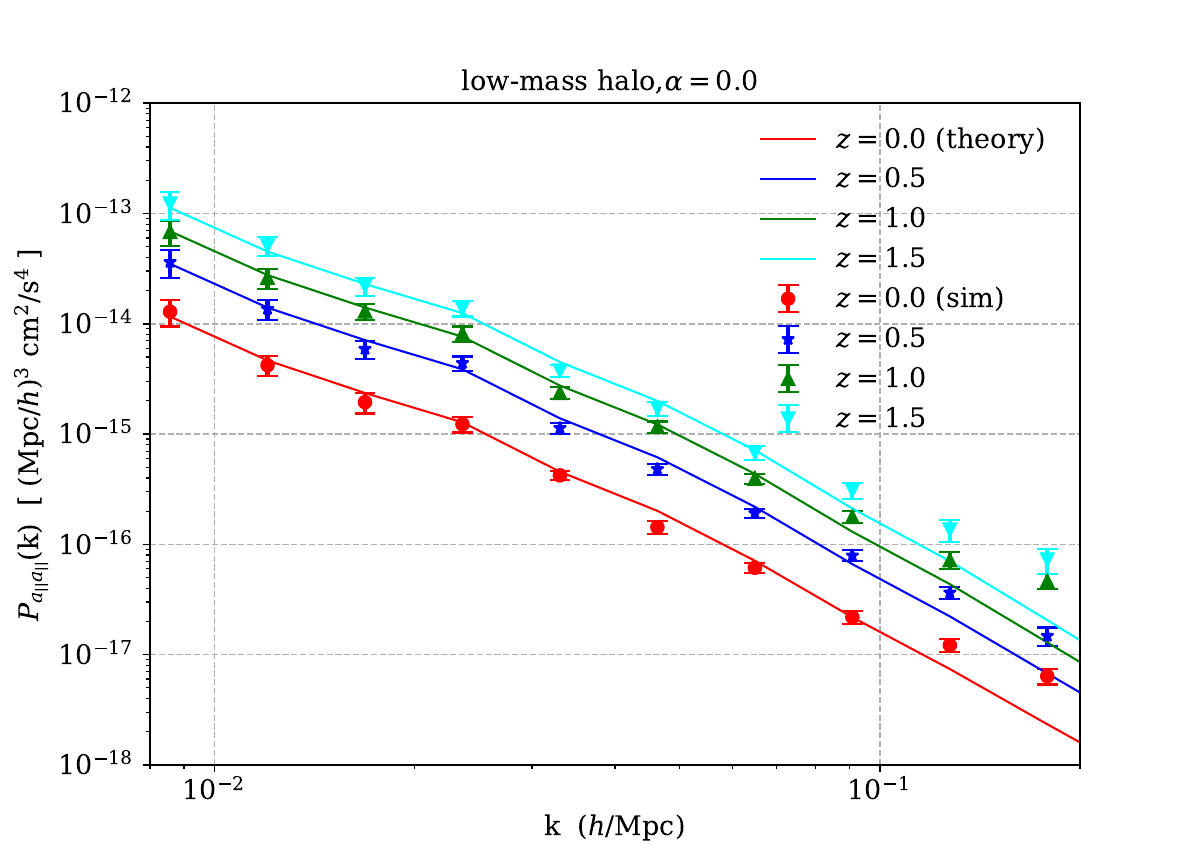}
    \includegraphics[width=0.45\linewidth]{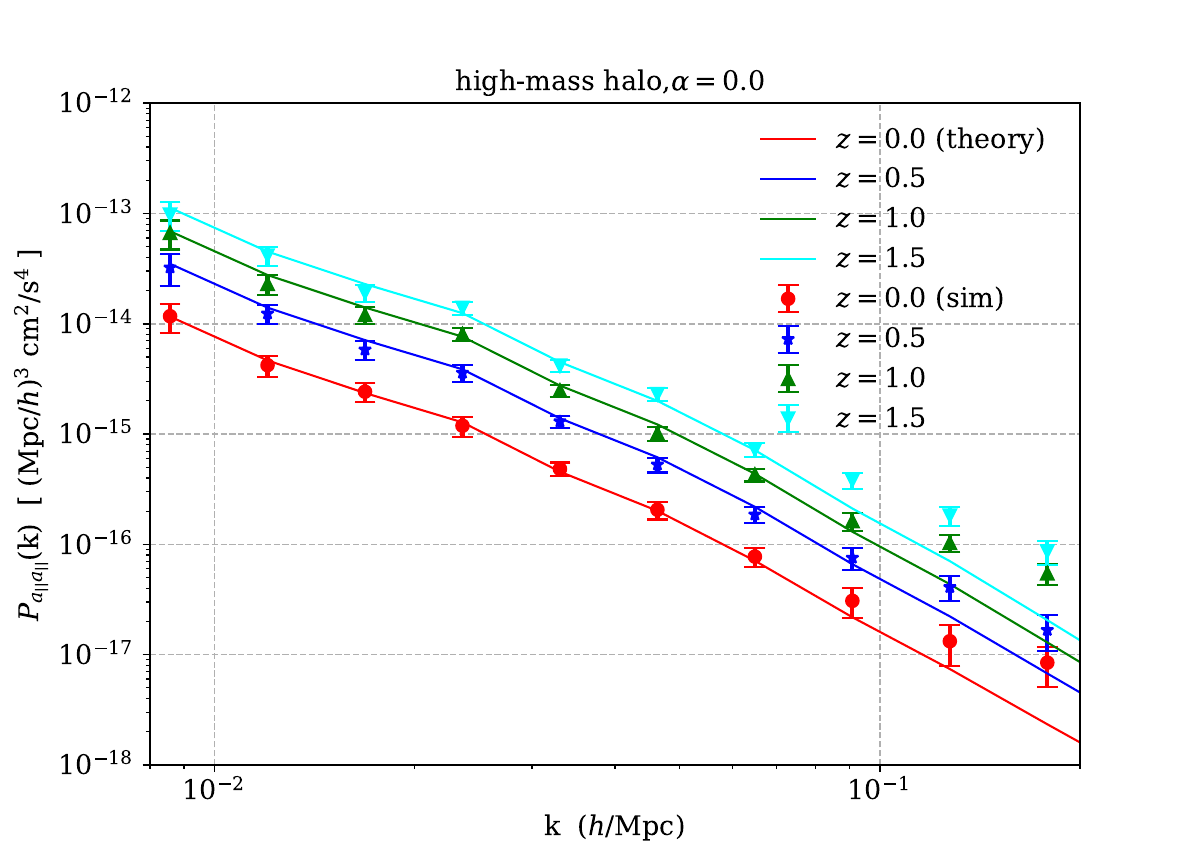}
    \caption{Comparison of halo acceleration power spectra from simulations and linear theory predictions for four redshift snapshots: small-mass halos (left) and large-mass halos (right). }
    \label{fig:Pa_halos}
\end{figure*}

The power spectra of the observed density field, $P_{\delta\delta}$, and the peculiar velocity field, $P_{vv}$, are related to the matter power spectrum, $P_m$, through the following relations:

\begin{equation}
P_{\delta \delta}=b^2 P_m \quad \text { and } \quad P_{v v}=\left(\frac{a H f}{k}\right)^2 P_m\,,
\end{equation}
where $a$ is the dimensionless scale factor, $H$ is the Hubble parameter, and $b$ is the linear bias factor. In a standard matter-dominated cosmology, the dimensionless growth rate $f$ is an analytical function of the matter density parameter $\Omega_m$, with a weak dependence on the cosmological constant $\Omega_{\Lambda}$. Similarly, within the framework of linear cosmological theory, the acceleration power spectrum is directly linked to the matter power spectrum.

As shown in Eq.~\ref{eq:a_pec}, the peculiar acceleration represents the integrated effect of excess gravitational forces caused by density fluctuations across the observable universe. Given a field of density perturbations, these accelerations can be directly computed using the Poisson equation. In comoving coordinates, the Poisson equation for the perturbed potential $\phi$ in a matter-dominated universe is: 
\begin{equation}
\nabla^2 \phi = \frac{3}{2} \Omega_m H^2 a^2 \delta_m\,,
\end{equation}
where $\delta_m$ is the matter density contrast. Transforming it into Fourier space yields:  
\begin{equation}
-k^2 \phi(\bm{k}, z) = \frac{3}{2} \Omega_m H^2 a^2 \delta_m\,.
\end{equation} 
From this, the linear relationship between the matter density contrast $\delta_m$ and the gravitational acceleration $\bm{a}_{\rm pec}$ can be derived. Therefore, the power spectrum of the peculiar gravitational acceleration is:  
\begin{equation}
P_{aa}(k, z) \equiv \left\langle \bm{a}_{\rm pec} \cdot \bm{a}_{\rm pec}^* \right\rangle = \left( \frac{3a H^2 \Omega_m}{2k} \right)^2 P_m(k, z)\,.
\end{equation}
The power spectrum of the peculiar acceleration along LOS is then:  
\begin{eqnarray} 
P_{a_\parallel a_\parallel}(k, z) &=& \frac{1}{3} P_{aa}(k, z) = \frac{3}{4} \left( \frac{a H^2 \Omega_m}{k} \right)^2 P_m(k, z)\nonumber\\
&=&\frac{3}{4} \left( \frac{ H^{2}_{0} \Omega_{m0}} {a^{2}k}   \right)^2 P_m(k, z)\,.
\end{eqnarray}

In Fig.~\ref{fig:Pa_halos}, we compare the power spectrum $P_{a_{\|} a_{\|}}$ with the corresponding measurements derived from low- and high-mass halo catalogs at four distinct redshifts. The shot noise contribution has been subtracted from the measured spectra. On linear scales ($k \lesssim 0.06~h/\rm Mpc$), the measurements exhibit excellent agreement with the predictions of linear theory for both halo mass bins. This agreement supports the earlier argument in the text that higher-order terms, such as $\delta_g a_{\mathrm{pec}, \|}$, can be neglected on linear scales. At smaller scales, however, the power spectra deviate from the linear predictions due to the influence of nonlinear evolution and halo assembly processes.

We also observe that, on scales \YZ{$k\lesssim 0.2~h/\rm Mpc$}, the power spectra for the two halo mass bins at the same redshift snapshot are nearly identical. This indicates that $E_{\rm ep}$ remains robust across this scale range. A more detailed discussion of this result is provided in Sect.~\ref{sec:res}.

\end{document}